\documentclass[english]{article}

\usepackage[T1]{fontenc}
\usepackage[latin9]{inputenc}
\usepackage{geometry}
\usepackage{amsmath}
\usepackage{setspace}
\usepackage{amssymb}
\usepackage{esint}

\makeatletter
\newcommand{\lyxaddress}[1]{
\par {\raggedright #1
\vspace{1.4em}
\noindent\par}
}

\usepackage{babel}
\makeatother

\begin{document}

\title{Generalized Gravi - Electromagnetism%
\thanks{\textbf{Oral presentation by Gaurav Karnatak  in, \textquoteleft{}Conference
on Gravitation Theories and Astronomy (CGTA- 2009)' on the eve of
International year of Astronomy at Nilkanthrao Shinde Science \& Arts
College, Bhadrawati, District-Chandrapur, Maharstra State, December
28-30, 2009 }%
}}

\author{P. S. Bisht, Gaurav Karnatak and O. P. S. Negi}

\maketitle

\lyxaddress{\begin{center}
Department of Physics,\\
Kumaun University,\\
S. S. J. Campus, \\
Almora-263601(Uttarakhand) India
\par\end{center}}

\lyxaddress{\begin{center}
Email- ps\_bisht123@rediffmail.com,\\
gauravkarnatak2009@yahoo.in,\\
ops\_negi@yahoo.com
\par\end{center}}

\begin{abstract}
A self consistant and manifestly covariant theory for the dynamics
of four charges (masses) (namely electric, magnetic, gravitational,
Heavisidian) has been developed in simple, compact and consistent
manner. Starting with an invariant Lagrangian density and its quaternionic
representation, we have obtained the consistent field equation for
the dynamics of four charges. It has been shown that the present reformulation
reproduces the dynamics of individual charges (masses) in the absence
of other charge (masses) as well as the generalized theory of dyons
(gravito - dyons) in the absence gravito - dyons (dyons).

\textbf{key words}: dyons, gravito - dyons, quaternion

\textbf{PACS NO}: 14.80Hv
\end{abstract}

\section{Introduction}

~~~~~~~~~Magnetic monopoles \cite{key-1} were advocated
to symmetrize Maxwell's equations in a manifest way that the mere
existence of an isolated magnetic charge implies the quantization
of electric charge. The fresh interests in this subject are enhanced
with the idea of t' Hooft and Polyakov \cite{key-2} that the classical
solutions having the properties of magnetic monopoles may be found
in Yang - Mills gauge theories. The Dirac monopoles were elementary
particles but the t' Hooft - Polyakov \cite{key-2} monopoles are
complicated extended object having a definite mass and finite size
inside of which massive fields play an important role in providing
a smooth structure and outside it they vanish rapidly leaving the
field configuration identical to abelian Dirac monopole. Julia and
Zee \cite{key-3} extended the theory of non-Abelian monopoles of
t' Hooft - Polyakov \cite{key-2} to the theory of non Abelian dyons
(particles carrying simultaneously electric and magnetic charges).
The quantum mechanical excitation of fundamental monopoles include
dyons which are automatically arisen from the semi-classical quantization
of global charge rotation degree of freedom of monopoles. Despite
of the potential importance of monopoles \cite{key-1,key-2} and dyons
\cite{key-3}, the formalism necessary to describe them has been clumsy
and not manifestly covariant. So, a self consistent and manifestly
covariant theory of generalized electromagnetic fields of dyons (particle
carrying electric and magnetic charges) and those for generalized
fields of gravitodyons, has been constructed \cite{key-4,key-5} in
terms of two four - potentials  to avoid the use of controversial
string variables. On the other hand,the quaternionic formulation \cite{key-6,key-7}
of electrodynamics has a long history \cite{key-8,key-9,key-10} ,
stretching back to Maxwell himself \cite{key-8} who used real (Hamilton)
quaternion in his original manuscript 'on the application of quaternion
to electromagnetism' and in his celebrated book {}``Treatise on Electricity
and Magnetism''. Quaternion analysis has since been rediscovered
at regular intervals and accordingly the Maxwell's Equations of electromagnetism
were rewritten as one quaternion equations \cite{key-11,key-12}.
Negi and coworkers \cite{key-13}, have also studied the quaternionic
formulation for generalized electromagnetic fields of dyons (particles
carrying simultaneous existence of electric and magnetic charges)
in unique, simpler and compact notations. Kravchenko and co-authors
\cite{key-14,key-15}, discussed the Maxwell's equations in homogeneous
media, chiral media and inhomogeneous media. Accordingly they have
developed the quaternionic reformulation of the time-dependent Maxwell's
equations along with the classical solution of a moving source i.e.
electron. In the series of papers Negi and coworkers \cite{key-16,key-17,key-18,key-19,key-20}
have derived the generalized Dirac - Maxwell (GDM) equations in presence
of electric and magnetic sources in an isotropic (homogeneous) medium.
They have also analyzed the other quantum equations of dyons in consistent
and manifest covariant way \cite{key-16}. This theory has been shown
to remain invariant under the duality transformations in isotropic
homogeneous medium. Quaternion analysis of time dependent Maxwell\textquoteright{}s
equations has been developed by them \cite{key-17} in presence of
electric and magnetic charges and the solutions for the classical
problem of moving charge (electric and magnetic) are obtained consistently.
The time dependent generalized Dirac - Maxwell\textquoteright{}s (GDM)
equations of dyons are also discussed \cite{key-18} in chiral and
inhomogeneous media and the solutions for the classical problem are
obtained. The quaternion reformulation of generalized electromagnetic
fields of dyons in chiral and inhomogeneous media has also been analyzed
\cite{key-19}. The monochromatic fields of generalized electromagnetic
fields of dyons have also discussed \cite{key-20} in slowly changing
media in a consistent manner. The quaternion analysis has also been
used \cite{key-21} to combine the complex desciption of dyons and
gravito - dyons and accordingly the unified quaternionic angular momentum
for generalized fields of dyons and gravito - dyons along with their
commutation relations has been analyzed in unique and consistent manner.
In this paper, we have made an attempt to developed a self consistant
and manifestly covariant theory for the dynamics of four charges (masses)
(namely electric, magnetic, gravitational, Heavisidian) has been developed
in simple, compact and consistent manner. Starting with an invariant
Lagrangian density and its quaternionic representation, we have obtained
the consistent field equation and equation of motion for the dynamics
of four charges. It has been shown that the present reformulation
reproduces the dynamics of individual charges (masses) in the absence
of other charge (masses) as well as the generalized theory of dyons
(gravito - dyons) in the absence gravito - dyons (dyons) and vice
versa..

\section{Quaternionic Unified Charge of Dyons and Gravito-dyons}

Let us describe the property of quaternion algebra with the use of
natural units $(c=\hbar=G=1)$ in order to reformulate the unified
theory of generalized electromagnetic fields (associated with dyons)
and generalized gravito - Heavisidian fields (associated with gravito
- dyons) of linear gravity. So we define the unified charge \cite{key-21,key-22}
as

\begin{eqnarray}
Q & =(e,\, g,\, m,\, h) & =e-ig-jm-kh,\label{eq:1}\end{eqnarray}
where $e,\, g,\, m$ and $h$ are respectively described as the electric,
magnetic, gravitational and Heavisidian charges (masses).In equation
(\ref{eq:1}), $i,\, j$ and $k$ are the quaternion units satisfy
the following properties

\begin{eqnarray}
ij & =-ji & =k\,(say)\label{eq:2}\end{eqnarray}
and

\begin{eqnarray}
i(ij) & =(ii)j & =-j,\nonumber \\
(ij)j & =i(jj) & =-i,\nonumber \\
ik & =-ki & =-j,\nonumber \\
kj & =-jk & =-i,\nonumber \\
i^{2}=j^{2}=k^{2} & = & -1.\label{eq:3}\end{eqnarray}
Complex structure $(e,\, g)$ represents the generalized charge of
dyons of electromagnetic fields while $(m,\, h)$ denotes the generalized
charge of gravito - dyons. The norm of unified quaternion charge (\ref{eq:1})
is 

\begin{eqnarray}
N(Q) & =Q\overline{Q} & =(e^{2}+g^{2}+m^{2}+h^{2}),\label{eq:4}\end{eqnarray}
where

\begin{eqnarray}
\overline{Q} & =(e,\,-g,\,-m,\,-h) & =e+ig+jm+kh.\label{eq:5}\end{eqnarray}
 Unified quaternion valued four-potential may then be defined as\begin{eqnarray}
\left\{ V_{\mu}\right\}  & = & \left\{ A_{\mu}\right\} -i\left\{ B_{\mu}\right\} -j\left\{ C_{\mu}\right\} -k\left\{ D_{\mu}\right\} ,\label{eq:6}\end{eqnarray}
where $\left\{ A_{\mu}\right\} $ is the four - potential associated
with the dynamics of electric charge, $\left\{ B_{\mu}\right\} $
is used for magnetic charge, $\left\{ C_{\mu}\right\} $ describes
the gravitational charge (mass) while $\left\{ D_{\mu}\right\} $
has been associated with the gravi - magnetic (Heavisidian) charge
(mass). Then the varius potentials of equation (\ref{eq:6}) are written
in the following quaternionic forms,

\begin{eqnarray}
A & = & A_{0}-iA_{1}-jA_{2}-kA_{3},\nonumber \\
B & = & B_{0}-iB_{1}-jB_{2}-kB_{3},\nonumber \\
C & = & C_{0}-iC_{1}-jC_{2}-kC_{3},\nonumber \\
D & = & D_{0}-iD_{1}-jD_{2}-kD_{3}.\label{eq:7}\end{eqnarray}
Similarly one may define \cite{key-22} the quaternion valued unified
field tensor as

\begin{eqnarray}
\Im_{\mu\nu} & = & F_{\mu\nu}-iM_{\mu\nu}-jf_{\mu\nu}-kN_{\mu\nu},\label{eq:8}\end{eqnarray}
where 

\begin{eqnarray}
F_{\mu\nu} & = & A_{\mu,\nu}-A_{\nu,\mu}-i\varepsilon_{\mu\nu\rho\sigma}B^{\rho\sigma},\nonumber \\
M_{\mu\nu} & = & B_{\mu,\nu}-B_{\nu,\mu}-i\varepsilon_{\mu\nu\rho\sigma}A^{\rho\sigma},\nonumber \\
f_{\mu\nu} & = & C_{\mu,\nu}-C_{\nu,\mu}-i\varepsilon_{\mu\nu\rho\sigma}D^{\rho\sigma},\nonumber \\
N_{\mu\nu} & = & D_{\mu,\nu}-D_{\nu,\mu}-i\varepsilon_{\mu\nu\rho\sigma}C^{\rho\sigma}\label{eq:9}\end{eqnarray}
are the field tensors respectively associated with the dynamics of
the electric, magnetic, gravitational and Heavisidian charges (masses).
These field tensors satisfy the following Maxwellian field equations\begin{eqnarray}
F_{\mu\nu,\nu} & = & j_{\mu}^{(e)},\nonumber \\
M_{\mu\nu,\nu} & = & j_{\mu}^{(m)},\nonumber \\
f_{\mu\nu,\nu} & = & j_{\mu}^{(G)},\nonumber \\
N_{\mu\nu,\nu} & = & j_{\mu}^{(H)},\label{eq:10}\end{eqnarray}
where $j_{\mu}^{(e)}$ is the four - current associated with electric
charge, $j_{\mu}^{(m)}$ is the four - current associated with magnetic
charge, $j_{\mu}^{(G)}$ is the four - current for gravitational charge(mass)
while $j_{\mu}^{(H)}$ is the four - current for Heavisidian charge
(mass). In equations (\ref{eq:8} - \ref{eq:10}), the field tensors
$F_{\mu\nu}$ and $M_{\mu\nu}$ are associated with dyons, while $f_{\mu\nu}$
and $N_{\mu\nu}$ are associated with gravito - dyons. As such the
quaternion valued current may then be defined as

\begin{eqnarray}
J_{\mu} & = & j_{\mu}^{(e)}-ij_{\mu}^{(m)}-jj_{\mu}^{(G)}-kj_{\mu}^{(H)}.\label{eq:11}\end{eqnarray}

\section{Lagrangian Formulation and field equations }

The Lagrangian density for the unified charged particle containing
the electric, magnetic, gravitational and Heavisidian charges and
the rest mass of the unified particle $M$, may be written in the
following form

\begin{align}
L= & -M-\frac{1}{4}\left[\alpha\left\{ (A_{\mu,\nu}-A_{\nu,\mu})^{2}-(B_{\mu,\nu}-B_{\nu,\mu})^{2}-(C_{\mu,\nu}-C_{\nu,\mu})^{2}-(D_{\mu,\nu}-D_{\nu,\mu})^{2}\right\} \right]\nonumber \\
- & 2\beta\left[\left(A_{\mu,\nu}-A_{\nu,\mu}\right)\left(B_{\mu,\nu}-B_{\nu,\mu}\right)+\left(C_{\mu,\nu}-C_{\nu,\mu}\right)\left(D_{\mu,\nu}-D_{\nu,\mu}\right)\right]\nonumber \\
- & 2\gamma\left[\left(A_{\mu,\nu}-A_{\nu,\mu}\right)\left(C_{\mu,\nu}-C_{\nu,\mu}\right)+\left(B_{\mu,\nu}-B_{\nu,\mu}\right)\left(D_{\mu,\nu}-D_{\nu,\mu}\right)\right]\nonumber \\
- & 2\Delta\left[\left(A_{\mu,\nu}-A_{\nu,\mu}\right)\left(D_{\mu,\nu}-D_{\nu,\mu}\right)+\left(B_{\mu,\nu}-B_{\nu,\mu}\right)\left(C_{\mu,\nu}-C_{\nu,\mu}\right)\right]\nonumber \\
+ & \left(\alpha A_{\mu}-\beta B_{\mu}+\gamma C_{\mu}-\Delta D_{\mu}\right)j_{\mu}^{(e)}-\left(\alpha B_{\mu}-\beta A_{\mu}+\gamma D_{\mu}-\Delta C_{\mu}\right)j_{\mu}^{(m)}\nonumber \\
+ & \left(\alpha C_{\mu}-\beta D_{\mu}+\gamma A_{\mu}-\Delta B_{\mu}\right)j_{\mu}^{(G)}-\left(\alpha D_{\mu}-\beta C_{\mu}+\gamma B_{\mu}-\Delta A_{\mu}\right)j_{\mu}^{(H)}\nonumber \\
= & L_{P}+L_{F}+L_{I},\label{eq:12}\end{align}
where $L_{P}$ is the free particle Lagrangian, $L_{F}$ is the field
Lagrangian and $L_{I}$ is the interaction Lagrangian and $\alpha,\,\beta,\,\gamma,\:\Delta$
are real positive arbitrary uni modular parameters satisfying the
following conditions

\begin{eqnarray}
\alpha-i\beta-j\gamma-k\Delta & = & e^{-\theta\hat{n}}=\cos\theta-\hat{n}\,\sin\theta,\label{eq:13}\end{eqnarray}
and

\begin{eqnarray}
\alpha+i\beta+j\gamma+k\Delta & = & e^{\theta\hat{n}}=\cos\theta+\hat{n}\,\sin\theta.\label{eq:14}\end{eqnarray}
From equations (\ref{eq:13}) and (\ref{eq:14}), we get

\begin{eqnarray}
\alpha^{2}+\beta^{2}+\gamma^{2}+\Delta^{2} & = & 1.\label{eq:15}\end{eqnarray}
As such, we may write the constancy condition \cite{key-13} as,

\begin{eqnarray}
\tan\theta= & \frac{g}{e}=\frac{h}{m}=\frac{B_{\mu}}{A_{\mu}} & =\frac{D_{\mu}}{C_{\mu}}=\frac{j_{\mu}^{(m)}}{j_{\mu}^{(e)}}=\frac{j_{\mu}^{(H)}}{j_{\mu}^{(G)}}.\label{eq:16}\end{eqnarray}
The action integral of the above unified system may be written as,

\begin{eqnarray}
S & = & \int Ldt=S_{P}+S_{F}+S_{I},\label{eq:17}\end{eqnarray}
where the action part $S_{P}$ depends upon the properties of the
particle, $S_{F}$ depends on the properties of the field and $S_{I}$
depends on the parameters of the particle and field both.

In deriving the equation of motion of the particle, we vary the trajectory
of the particle without changing the field parameters and as such
the action $S_{F}$ does not affect the motion. On the other hand,in
order to find the field equations we take the variation with respect
to field parameters assuming the trajectory of the particle fixed.
For deriving the equation of motion, we write the concerned part of
action in the following form,

\begin{align}
S=S_{P}+S_{I} & =-\intop_{_{t_{1}}}^{t_{2}}Mdt\nonumber \\
+ & \intop_{_{t_{1}}}^{t_{2}}\left(\alpha A_{\mu}-\beta B_{\mu}+\gamma C_{\mu}-\Delta D_{\mu}\right)j_{0}^{(e)}\frac{dx_{\mu}}{dt}dt\nonumber \\
- & \intop_{_{t_{1}}}^{t_{2}}\left(\alpha B_{\mu}-\beta A_{\mu}+\gamma D_{\mu}-\Delta C_{\mu}\right)j_{0}^{(m)}\frac{dx_{\mu}}{dt}dt\nonumber \\
+ & \intop_{_{t_{1}}}^{t_{2}}\left(\alpha C_{\mu}-\beta D_{\mu}+\gamma A_{\mu}-\Delta B_{\mu}\right)j_{0}^{(G)}\frac{dx_{\mu}}{dt}dt\nonumber \\
- & \intop_{_{t_{1}}}^{t_{2}}\left(\alpha D_{\mu}-\beta C_{\mu}+\gamma B_{\mu}-\Delta A_{\mu}\right)j_{0}^{(H)}\frac{dx_{\mu}}{dt}dt,\label{eq:18}\end{align}
where $j_{\mu}^{(e)}$, $j_{\mu}^{(m)}$, $j_{\mu}^{(G)}$ and $j_{\mu}^{(H)}$
are expressed in terms of $j_{0}^{(e)}$, $j_{0}^{(m)}$, $j_{0}^{(G)}$
and $j_{0}^{(H)}$ and $\frac{dx_{\mu}}{dt}$. Equation (\ref{eq:18})
may also be written as,

\begin{align}
S & =-M\int_{a}^{^{b}}dS\nonumber \\
+ & \int_{a}^{^{b}}[\left(\alpha A_{\mu}-\beta B_{\mu}+\gamma C_{\mu}-\Delta D_{\mu}\right)j_{0}^{(e)}dx_{\mu}\nonumber \\
- & \int_{a}^{b}\left(\alpha B_{\mu}-\beta A_{\mu}+\gamma D_{\mu}-\Delta C_{\mu}\right)j_{0}^{(m)}dx_{\mu}\nonumber \\
+ & \int_{a}^{b}\left(\alpha C_{\mu}-\beta D_{\mu}+\gamma A_{\mu}-\Delta B_{\mu}\right)j_{0}^{(G)}dx_{\mu}\nonumber \\
- & \int_{a}^{b}\left(\alpha D_{\mu}-\beta C_{\mu}+\gamma B_{\mu}-\Delta A_{\mu}\right)j_{0}^{(H)}dx_{\mu},\label{eq:19}\end{align}
where the first term is an integral along the world line of the particle
between two events $a$ and $b$ i.e the presence of the particle
at its initial and final positions at time $t_{1}$ and $t_{2}$.
The variation of the action $S$ may be written as,

\begin{align}
\delta S & =\delta\{\int_{a}^{^{b}}-MdS\nonumber \\
+ & \int_{a}^{^{b}}[\left(\alpha A_{\mu}-\beta B_{\mu}+\gamma C_{\mu}-\Delta D_{\mu}\right)j_{0}^{(e)}dx_{\mu}\nonumber \\
- & \left(\alpha B_{\mu}-\beta A_{\mu}+\gamma D_{\mu}-\Delta C_{\mu}\right)j_{0}^{(m)}dx_{\mu}\nonumber \\
+ & \left(\alpha C_{\mu}-\beta D_{\mu}+\gamma A_{\mu}-\Delta B_{\mu}\right)j_{0}^{(G)}dx_{\mu}\nonumber \\
- & \left(\alpha D_{\mu}-\beta C_{\mu}+\gamma B_{\mu}-\Delta A_{\mu}\right)j_{0}^{(H)}dx_{\mu}]\}=0.\label{eq:20}\end{align}
Taking the variation of the terms one by one, we get

\begin{eqnarray}
I & = & \delta\int_{a}^{^{b}}-MdS=\int_{a}^{^{b}}Mu_{\mu}d\delta x_{\mu},\label{eq:21}\end{eqnarray}
where $u_{\mu}=\frac{dx_{\mu}}{dS}$ (four - velocity). Integrationg
equation (\ref{eq:21}) by parts, we get 

\begin{eqnarray}
I & = & \left|Mu_{\mu}\delta x_{\mu}\right|_{a}^{b}-\int_{a}^{^{b}}Mdu_{\mu}\delta x_{\mu},\label{eq:22}\end{eqnarray}
where first term is zero, since the integral is varied between fixed
limits,

\begin{eqnarray}
(\delta x_{\mu})_{b} & = & (\delta x_{\mu})_{a}=0.\label{eq:23}\end{eqnarray}
Thus

\begin{eqnarray}
I= & -\int_{a}^{^{b}}Mdu_{\mu}\delta x_{\mu} & =-\int_{a}^{^{b}}M\,\frac{du_{\mu}}{dS}\, dS\,\delta x_{\mu}.\label{eq:24}\end{eqnarray}
Now. it is convenient to take the following variations in order to
solve the equations of motion of varius charges,

\begin{align}
\delta\int_{a}^{^{b}}\alpha A_{\mu}j_{0}^{(e)}dx_{\mu} & =\int_{a}^{^{b}}\alpha j_{0}^{(e)}(A_{\nu,\mu}-A_{\mu,\nu})u_{\nu}\delta x_{\mu}dS=\int_{a}^{^{b}}\alpha j_{0}^{(e)}F_{\mu\nu}u_{\nu}\delta x_{\mu}dS;\nonumber \\
\delta\int_{a}^{^{b}}\beta B_{\mu}j_{0}^{(e)}dx_{\mu} & =\int_{a}^{^{b}}\beta j_{0}^{(e)}(B_{\nu,\mu}-B_{\mu,\nu})u_{\nu}\delta x_{\mu}dS=\int_{a}^{^{b}}\beta j_{0}^{(e)}M_{\mu\nu}u_{\nu}\delta x_{\mu}dS;\nonumber \\
\delta\int_{a}^{^{b}}\gamma C_{\mu}j_{0}^{(e)}dx_{\mu} & =\int_{a}^{^{b}}\gamma j_{0}^{(e)}(C_{\nu,\mu}-C_{\mu,\nu})u_{\nu}\delta x_{\mu}dS=\int_{a}^{^{b}}\gamma j_{0}^{(e)}f_{\mu\nu}u_{\nu}\delta x_{\mu}dS;\nonumber \\
\delta\int_{a}^{^{b}}\Delta D_{\mu}j_{0}^{(e)}dx_{\mu} & =\int_{a}^{^{b}}\Delta j_{0}^{(e)}(D_{\nu,\mu}-D_{\mu,\nu})u_{\nu}\delta x_{\mu}dS=\int_{a}^{^{b}}\Delta j_{0}^{(e)}N_{\mu\nu}u_{\nu}\delta x_{\mu}dS;\label{eq:25}\end{align}

\begin{align}
\delta\int_{a}^{^{b}}\alpha B_{\mu}j_{0}^{(m)}dx_{\mu} & =\int_{a}^{^{b}}\alpha j_{0}^{(m)}(B_{\nu,\mu}-B_{\mu,\nu})u_{\nu}\delta x_{\mu}dS=\int_{a}^{^{b}}\alpha j_{0}^{(m)}M_{\mu\nu}u_{\nu}\delta x_{\mu}dS;\nonumber \\
\delta\int_{a}^{^{b}}\beta A_{\mu}j_{0}^{(m)}dx_{\mu} & =\int_{a}^{^{b}}\beta j_{0}^{(m)}(A_{\nu,\mu}-A_{\mu,\nu})u_{\nu}\delta x_{\mu}dS=\int_{a}^{^{b}}\beta j_{0}^{(m)}F_{\mu\nu}u_{\nu}\delta x_{\mu}dS;\nonumber \\
\delta\int_{a}^{^{b}}\gamma D_{\mu}j_{0}^{(m)}dx_{\mu} & =\int_{a}^{^{b}}\gamma j_{0}^{(m)}(D_{\nu,\mu}-D_{\mu,\nu})u_{\nu}\delta x_{\mu}dS=\int_{a}^{^{b}}\gamma j_{0}^{(m)}N_{\mu\nu}u_{\nu}\delta x_{\mu}dS;\nonumber \\
\delta\int_{a}^{^{b}}\Delta C_{\mu}j_{0}^{(m)}dx_{\mu} & =\int_{a}^{^{b}}\Delta j_{0}^{(m)}(C_{\nu,\mu}-C_{\mu,\nu})u_{\nu}\delta x_{\mu}dS=\int_{a}^{^{b}}\Delta j_{0}^{(m)}f_{\mu\nu}u_{\nu}\delta x_{\mu}dS;\label{eq:26}\end{align}

\begin{align}
\delta\int_{a}^{^{b}}\alpha C_{\mu}j_{0}^{(G)}dx_{\mu} & =\int_{a}^{^{b}}\alpha j_{0}^{(G)}(C_{\nu,\mu}-C_{\mu,\nu})u_{\nu}\delta x_{\mu}dS=\int_{a}^{^{b}}\alpha j_{0}^{(G)}f_{\mu\nu}u_{\nu}\delta x_{\mu}dS;\nonumber \\
\delta\int_{a}^{^{b}}\beta D_{\mu}j_{0}^{(G)}dx_{\mu} & =\int_{a}^{^{b}}\beta j_{0}^{(G)}(D_{\nu,\mu}-D_{\mu,\nu})u_{\nu}\delta x_{\mu}dS=\int_{a}^{^{b}}\beta j_{0}^{(G)}N_{\mu\nu}u_{\nu}\delta x_{\mu}dS;\nonumber \\
\delta\int_{a}^{^{b}}\gamma A_{\mu}j_{0}^{(G)}dx_{\mu} & =\int_{a}^{^{b}}\gamma j_{0}^{(G)}(A_{\nu,\mu}-A_{\mu,\nu})u_{\nu}\delta x_{\mu}dS=\int_{a}^{^{b}}\gamma j_{0}^{(G)}F_{\mu\nu}u_{\nu}\delta x_{\mu}dS;\nonumber \\
\delta\int_{a}^{^{b}}\Delta B_{\mu}j_{0}^{(G)}dx_{\mu} & =\int_{a}^{^{b}}\Delta j_{0}^{(G)}(B_{\nu,\mu}-B_{\mu,\nu})u_{\nu}\delta x_{\mu}dS=\int_{a}^{^{b}}\Delta j_{0}^{(G)}M_{\mu\nu}u_{\nu}\delta x_{\mu}dS;\label{eq:27}\end{align}
and

\begin{align}
\delta\int_{a}^{^{b}}\alpha D_{\mu}j_{0}^{(H)}dx_{\mu} & =\int_{a}^{^{b}}\alpha j_{0}^{(H)}(D_{\nu,\mu}-D_{\mu,\nu})u_{\nu}\delta x_{\mu}dS=\int_{a}^{^{b}}\alpha j_{0}^{(H)}N_{\mu\nu}u_{\nu}\delta x_{\mu}dS;\nonumber \\
\delta\int_{a}^{^{b}}\beta C_{\mu}j_{0}^{(H)}dx_{\mu} & =\int_{a}^{^{b}}\beta j_{0}^{(H)}(C_{\nu,\mu}-C_{\mu,\nu})u_{\nu}\delta x_{\mu}dS=\int_{a}^{^{b}}\beta j_{0}^{(H)}f_{\mu\nu}u_{\nu}\delta x_{\mu}dS;\nonumber \\
\delta\int_{a}^{^{b}}\gamma A_{\mu}j_{0}^{(H)}dx_{\mu} & =\int_{a}^{^{b}}\gamma j_{0}^{(H)}(A_{\nu,\mu}-A_{\mu,\nu})u_{\nu}\delta x_{\mu}dS=\int_{a}^{^{b}}\gamma j_{0}^{(H)}F_{\mu\nu}u_{\nu}\delta x_{\mu}dS;\nonumber \\
\delta\int_{a}^{^{b}}\Delta B_{\mu}j_{0}^{(H)}dx_{\mu} & =\int_{a}^{^{b}}\Delta j_{0}^{(H)}(B_{\nu,\mu}-B_{\mu,\nu})u_{\nu}\delta x_{\mu}dS=\int_{a}^{^{b}}\Delta j_{0}^{(H)}M_{\mu\nu}u_{\nu}\delta x_{\mu}dS.\label{eq:28}\end{align}
 Substituting equations (\ref{eq:24}-\ref{eq:28}) in equation (\ref{eq:20}),
we get the following equations of motion for the dynamics of fields
of a particle containing four charges as,

\begin{align}
M\,\frac{d^{2}x_{\mu}}{dt^{2}} & =e\, F_{\mu\nu}u^{\nu};\nonumber \\
M\,\frac{d^{2}x_{\mu}}{dt^{2}} & =g\, M_{\mu\nu}u^{\nu};\nonumber \\
M\,\frac{d^{2}x_{\mu}}{dt^{2}} & =m\, f_{\mu\nu}u^{\nu};\nonumber \\
M\,\frac{d^{2}x_{\mu}}{dt^{2}} & =h\, N_{\mu\nu}u^{\nu}.\label{eq:29}\end{align}
Here $M$ is the effective mass given by 

\begin{align}
M= & m\,-\frac{\kappa-1}{2}\, h\label{eq:30}\end{align}
where $\kappa=+1$ for elcrtomagnetic fields and $\kappa=-1$for gravito-Heavisidian
fields. The field equations (\ref{eq:29}) may also be obtained from
the action (\ref{eq:17}) by taking the trajectory of the unified
particle fixed and considering the variation of the field parameters
(i.e. potentials) only. So, the parameters associated unified charged
particle such as the free particle action and four - current density
$J_{\mu}$ are treated to be constant i.e. $\delta S_{P}=0.$ From
equation (\ref{eq:12}), we may write the field and interaction parts
of the Lagrangian required for the variation as,

\begin{align}
L_{F}+L_{I} & =-\frac{1}{4}\,\left[\alpha(F_{\mu\nu}F^{\mu\nu}-M_{\mu\nu}M^{\mu\nu}-f_{\mu\nu}f^{\mu\nu}-N_{\mu\nu}N^{\mu\nu})\right]\nonumber \\
-2\beta\, & \left[F_{\mu\nu}M_{\mu\nu}+f_{\mu\nu}N_{\mu\nu}\right]-2\gamma\,\left[F_{\mu\nu}f_{\mu\nu}+M_{\mu\nu}N_{\mu\nu}\right]-2\Delta\left[F_{\mu\nu}N_{\mu\nu}+f_{\mu\nu}M_{\mu\nu}\right]\nonumber \\
+ & \left(\alpha A_{\mu}-\beta B_{\mu}+\gamma C_{\mu}-\Delta D_{\mu}\right)j_{\mu}^{(e)}-\left(\alpha B_{\mu}-\beta A_{\mu}+\gamma D_{\mu}-\Delta C_{\mu}\right)j_{\mu}^{(m)}\nonumber \\
+ & \left(\alpha C_{\mu}-\beta D_{\mu}+\gamma A_{\mu}-\Delta B_{\mu}\right)j_{\mu}^{(G)}-\left(\alpha D_{\mu}-\beta C_{\mu}+\gamma B_{\mu}-\Delta A_{\mu}\right)j_{\mu}^{(H)}.\label{eq:31}\end{align}
Now considering the term wise variation in equation (\ref{eq:31}),
we get

\begin{align}
\delta S_{1} & =\int\frac{-\alpha}{4}\delta F_{\mu\nu}^{2}d\Omega=\int\frac{-\alpha}{2}F_{\mu\nu}\delta F_{\mu\nu}d\Omega;\nonumber \\
\delta S_{2} & =\int\frac{-\alpha}{4}\delta M_{\mu\nu}^{2}d\Omega=\int\frac{-\alpha}{2}M_{\mu\nu}\delta M_{\mu\nu}d\Omega;\nonumber \\
\delta S_{3} & =\int\frac{-\alpha}{4}\delta f_{\mu\nu}^{2}d\Omega=\int\frac{-\alpha}{2}f_{\mu\nu}\delta f_{\mu\nu}d\Omega;\nonumber \\
\delta S_{4} & =\int\frac{-\alpha}{4}\delta N_{\mu\nu}^{2}d\Omega=\int\frac{-\alpha}{2}N_{\mu\nu}\delta N_{\mu\nu}d\Omega;\label{eq:32}\end{align}

\begin{align}
\delta S_{5} & =\int\frac{\beta}{2}\delta(F_{\mu\nu}M_{\mu\nu}+f_{\mu\nu}N_{\mu\nu})d\Omega\nonumber \\
= & \int\frac{\beta}{2}\left\{ \left[\, F_{\mu\nu}\delta M_{\mu\nu}+M_{\mu\nu}\delta F_{\mu\nu}\right]+\left[\, f_{\mu\nu}\delta N_{\mu\nu}+N_{\mu\nu}\delta f_{\mu\nu}\right]\right\} d\Omega;\label{eq:33}\end{align}

\begin{align}
\delta S_{6} & =\int\frac{\gamma}{2}\delta(F_{\mu\nu}f_{\mu\nu}+M_{\mu\nu}N_{\mu\nu})d\Omega\nonumber \\
= & \int\frac{\gamma}{2}\left\{ \left[\, F_{\mu\nu}\delta f_{\mu\nu}+f_{\mu\nu}\delta F_{\mu\nu}\right]+\left[\, M_{\mu\nu}\delta N_{\mu\nu}+N_{\mu\nu}\delta M_{\mu\nu}\right]\right\} d\Omega;\label{eq:34}\end{align}

\begin{align}
\delta S_{7} & =\int\frac{\Delta}{2}\delta(F_{\mu\nu}N_{\mu\nu}+M_{\mu\nu}f_{\mu\nu})d\Omega\nonumber \\
= & \int\frac{\Delta}{2}\left\{ \left[\, F_{\mu\nu}\delta N_{\mu\nu}+N_{\mu\nu}\delta F_{\mu\nu}\right]+\left[\, M_{\mu\nu}\delta f_{\mu\nu}+f_{\mu\nu}\delta M_{\mu\nu}\right]\right\} d\Omega;\label{eq:35}\end{align}

\begin{align}
\delta S_{8} & =\int\alpha j_{\mu}^{(e)}\delta A_{\mu}-\int\beta j_{\mu}^{(e)}\delta B_{\mu}+\int\gamma j_{\mu}^{(e)}\delta C_{\mu}-\int\Delta j_{\mu}^{(e)}\delta D_{\mu}\nonumber \\
- & \int\alpha j_{\mu}^{(m)}\delta B_{\mu}+\int\beta j_{\mu}^{(m)}\delta A_{\mu}-\int\gamma j_{\mu}^{(m)}\delta D_{\mu}+\int\triangle j_{\mu}^{(m)}\delta C_{\mu}\nonumber \\
+ & \int\alpha j_{\mu}^{(G)}\delta C_{\mu}-\int\beta j_{\mu}^{(G)}\delta D_{\mu}+\int\gamma j_{\mu}^{(G)}\delta A_{\mu}-\int\Delta j_{\mu}^{(G)}\delta D_{\mu}\nonumber \\
- & \int\alpha j_{\mu}^{(H)}\delta D_{\mu}+\int\beta j_{\mu}^{(H)}\delta C_{\mu}-\int\gamma j_{\mu}^{(H)}\delta B_{\mu}+\int\triangle j_{\mu}^{(H)}\delta A_{\mu}.\label{eq:36}\end{align}
$\delta S_{1}$ may be calculated after integrating it by parts and
applying Gauss theorem as 

\begin{eqnarray}
\delta S_{1} & = & -\alpha\int\frac{\partial F_{\mu\nu}}{\partial x_{\mu}}\delta A_{\mu}d\Omega.\label{eq:37}\end{eqnarray}
Similarly, other variations may be calculated. taking into account
the equation

\begin{equation}
\delta S=\delta S_{1}+\delta S_{2}+\delta S_{3}+\delta S_{4}+\delta S_{5}+\delta S_{6}+\delta S_{7}+\delta S_{8}=0\label{eq:38}\end{equation}
from which we get

\begin{align}
\int & \left(j_{\mu}^{(e)}\delta A_{\mu}+j_{\mu}^{(m)}\delta B_{\mu}+j_{\mu}^{(G)}\delta C_{\mu}+j_{\mu}^{(H)}\delta D_{\mu}\right)\alpha\, d\Omega\nonumber \\
- & \int\left(\frac{\partial F_{\mu\nu}}{\partial x_{\mu}}\delta A_{\mu}-\frac{\partial M_{\mu\nu}}{\partial x_{\mu}}\delta B_{\mu}-\frac{\partial f_{\mu\nu}}{\partial x_{\mu}}\delta C_{\mu}-\frac{\partial N_{\mu\nu}}{\partial x_{\mu}}\delta D_{\mu}\right)\alpha\, d\Omega\nonumber \\
- & \int\left(j_{\mu}^{(e)}\delta B_{\mu}+j_{\mu}^{(m)}\delta A_{\mu}+j_{\mu}^{(G)}\delta D_{\mu}+j_{\mu}^{(H)}\delta C_{\mu}\right)\alpha\, d\Omega\nonumber \\
- & \int\left(\frac{\partial F_{\mu\nu}}{\partial x_{\mu}}\delta A_{\mu}-\frac{\partial M_{\mu\nu}}{\partial x_{\mu}}\delta B_{\mu}-\frac{\partial f_{\mu\nu}}{\partial x_{\mu}}\delta C_{\mu}-\frac{\partial N_{\mu\nu}}{\partial x_{\mu}}\delta D_{\mu}\right)\alpha\, d\Omega|\nonumber \\
- & \int(j_{\mu}^{(e)}-ij_{\mu}^{(m)}-jj_{\mu}^{(G)}-kj_{\mu}^{(H)})\delta(A_{\mu}+iB_{\mu}+jC_{\mu}+kD_{\mu})\beta\, d\Omega\nonumber \\
- & \int(F_{\mu\nu,\nu}-iM_{\mu\nu,\nu}-jf_{\mu\nu,\nu}-kN_{\mu\nu,\nu})\delta(A_{\mu}+iB_{\mu}+jC_{\mu}+D_{\mu})\beta\, d\Omega\nonumber \\
- & \int(j_{\mu}^{(e)}-ij_{\mu}^{(m)}-jj_{\mu}^{(G)}-kj_{\mu}^{(H)})\delta(B_{\mu}+iA_{\mu}+jD_{\mu}+kC_{\mu})\beta\, d\Omega\nonumber \\
- & (F_{\mu\nu,\nu}-iM_{\mu\nu,\nu}-jf_{\mu\nu,\nu}-kN_{\mu\nu,\nu})\delta(B_{\mu}+iA_{\mu}+jD_{\mu}+kC_{\mu})\beta\, d\Omega\nonumber \\
= & 0\label{eq:39}\end{align}
which yields the Maxwellian field equations given by equation (\ref{eq:10})
for the dynamics of electric, magnetic, gravitational and Heavisidian
charges (masses). These equations thus provide the following form
of unified field equations of a particle simultaneously contains these
four charges (masses) 

\begin{eqnarray}
\Im_{\mu\nu,\nu} & = & J_{\mu}\label{eq:40}\end{eqnarray}
where $\Im_{\mu\nu}$ is the unified field tensor defind by equation
(\ref{eq:8}) and $J_{\mu}$is the unified current density given by
equation (\ref{eq:11}).

\section{Equation of motion of a unified charged in quaternionic form }

From equation (\ref{eq:26}), we may write the unified equation of
motion for a particle simultaneously contains four charges (masses)
in terms of quaternion as 

\begin{eqnarray}
M\frac{d^{2}x_{\mu}}{dt^{2}} & = & \left\{ Re\left(\bar{Q}\Im_{\mu\nu}\right)\right\} u^{\nu}\label{eq:41}\end{eqnarray}
where $\ddot{x_{\mu}}=\frac{d^{2}x_{\mu}}{dt^{2}}$ is particle acceleration,
$u^{\nu}$ is the four - velocity, and $\overline{Q}$ is the quaternion
conjugate of the unified charge given by equation (\ref{eq:5}). The
right hand side of the equation (\ref{eq:41}) may also be written
as 

\begin{eqnarray}
\left\{ Re\left(\bar{Q}\Im_{\mu\nu}\right)\right\} u^{\nu} & = & \left(eF_{\mu\nu}+gM_{\mu\nu}+mf_{\mu\nu}+hN_{\mu\nu}\right)u^{\nu}\label{eq:42}\end{eqnarray}
which can be reduced as 

\begin{align}
Re\left[Q,\,\Im_{\mu\nu}\right]u^{\nu} & =\frac{1}{2}\left(Q\bar{\Im}_{\mu\nu}+\bar{Q}\Im_{\mu\nu}\right)u^{\nu}\nonumber \\
= & \left(eF_{\mu\nu}+gM_{\mu\nu}+mf_{\mu\nu}+hN_{\mu\nu}\right)u^{\nu}\nonumber \\
= & \left\{ e\left[\overrightarrow{E}+\overrightarrow{v}\times\overrightarrow{H}\right]+g\left[\overrightarrow{H}-\overrightarrow{v}\times\overrightarrow{E}\right]\right\} \nonumber \\
+ & \left\{ m\left[\overrightarrow{G}-\overrightarrow{v}\times\overrightarrow{\mathcal{H}}\right]+h\left[\overrightarrow{\mathcal{H}}+\overrightarrow{v}\times\overrightarrow{G}\right]\right\} \label{eq:43}\end{align}
where $\overrightarrow{E}$ is the electric field, $\overrightarrow{H}$
is the magnetic field, $\overrightarrow{G}$ is the gravitational
field and $\overrightarrow{\mathcal{H}}$ is Heavisidian field.

\section{Euler's Lagrangian equation of motion of unified charged }

The Lagrangian density for unified fields of a paricle containing
simultaneously the four charges namely electric, magnetic, gravitational
and Heavisidian may also be expressed as 

\begin{equation}
L=-\frac{1}{4}F_{\mu\nu}F^{\mu\nu}-\frac{1}{4}M_{\mu\nu}M^{\mu\nu}-\frac{1}{4}f_{\mu\nu}f^{\mu\nu}-\frac{1}{4}N_{\mu\nu}N^{\mu\nu}+A_{\mu}j_{\mu}^{(e)}+B_{\mu}j_{\mu}^{(m)}+C_{\mu}j_{\mu}^{(G)}+D_{\mu}j_{\mu}^{(H)}.\label{eq:44}\end{equation}
This Lagrangian density may also be written in terms of Grassmann
product as 

\begin{eqnarray}
L & = & =\frac{1}{8}\left[\Im_{\mu\nu},\overline{\Im}_{\mu\nu}\right]+\left[V_{\mu},\overline{J}_{\mu}\right]\label{eq:45}\end{eqnarray}
where $\left[V_{\mu},\overline{J}_{\mu}\right]=\frac{1}{2}\left[V_{\mu}\overline{J}_{\mu}+\overline{V}_{\mu}J_{\mu}\right]$
, with $\overline{V}_{\mu}$ (the quaternion conjugate of the unified
four - potential), $\overline{J}_{\mu}$ (the quaternion conjugate
of unified four - current density) and $\overline{\Im}_{\mu\nu}$
(the quaternion conjugate of unified field tensor) are defined as
follows,

\begin{eqnarray}
\overline{V}_{\mu} & = & A_{\mu}+iB_{\mu}+jC_{\mu}+kD_{\mu};\nonumber \\
\overline{J}_{\mu} & = & j_{\mu}^{(e)}+ij_{\mu}^{(m)}+jj_{\mu}^{(G)}+kj_{\mu}^{(H)};\nonumber \\
\overline{\Im}_{\mu\nu} & = & F_{\mu\nu}+iM_{\mu\nu}+jf_{\mu\nu}+kN_{\mu\nu}.\label{eq:46}\end{eqnarray}
Then

\begin{eqnarray}
\left[V_{\mu},\overline{J}_{\mu}\right] & = & \left[A_{\mu}j_{\mu}^{(e)}+B_{\mu}j_{\mu}^{(m)}+C_{\mu}j_{\mu}^{(G)}+D_{\mu}j_{\mu}^{(H)}\right].\label{eq:47}\end{eqnarray}
Similarly, we have 

\begin{eqnarray}
\frac{1}{8}\left[\Im_{\mu\nu},\overline{\Im}_{\mu\nu}\right] & = & \frac{1}{4}\left[F_{\mu\nu}F^{\mu\nu}+M_{\mu\nu}M^{\mu\nu}+f_{\mu\nu}f^{\mu\nu}+N_{\mu\nu}N^{\mu\nu}\right].\label{eq:48}\end{eqnarray}
From the Lagrangian density (\ref{eq:44}), we get the Euler's Lagrangian
equations in the following forms;

\begin{eqnarray}
\frac{\partial L}{\partial A_{\mu}}-\partial_{\nu}\frac{\partial L}{\partial(\partial_{\nu}A_{\mu})} & = & 0;\nonumber \\
\frac{\partial L}{\partial B_{\mu}}-\partial_{\nu}\frac{\partial L}{\partial(\partial_{\nu}B_{\mu})} & = & 0;\nonumber \\
\frac{\partial L}{\partial C_{\mu}}-\partial_{\nu}\frac{\partial L}{\partial(\partial_{\nu}C_{\mu})} & = & 0;\nonumber \\
\frac{\partial L}{\partial D_{\mu}}-\partial_{\nu}\frac{\partial L}{\partial(\partial_{\nu}D_{\mu})} & = & 0.\label{eq:49}\end{eqnarray}
This equation provides the field equations associated respectively
with the dynamics of electric, magnetic, gravitational and Heavisidian
charges (masses) given by equation (\ref{eq:10}) after taking care
the usual method of variations with respect to potential. These equation
may immediately be genralized to equation (\ref{eq:40}) as unified
field equations of a particle simultaneously containing four charges
namely electric, magnetic, gravitational and Heavisidian charges (masses).

\section{Conclusion}

A unified Lagrangian density of generalized electromagnetic fields
and Heavisidian fields have been developed with the fact that these
fields essentially possess the structural symmetry. The action integral
(\ref{eq:17}) of the unified quaternionic charge (i.e. the complex
structure of dyons and gravito - dyons) in the field of others constructed
by choosing the suitable Lagrangian density (\ref{eq:12}), does not
make the use of string variables. Rather, it has been written in terms
of four - potentials associated with four charges namely electric,
magnetic, gravitational and Heavisidian charges. Following the usual
method of variation, it is shown that this action leads to equation
of motion (\ref{eq:29}) and unified field equations (\ref{eq:40}).
The beauty of the Lagrangian (\ref{eq:12}) lies in the fact that
individual components of four - currents and four - potentials have
been embodied in the corresponding generalized quantities through
the unknown parameters $\alpha,\,\beta,\,\gamma$ and $\Delta$ which
satisfy the constraints described by equations (\ref{eq:13}, \ref{eq:14})
and (\ref{eq:15}). It has already been shown that the Lagrangian
density (\ref{eq:12}), equation of motion (\ref{eq:29}) and the
unified field equations (\ref{eq:40}) are invariant under the rotation
in charge space or its combination with space and time reflections
and also under the reflection in charge space combined with time reversal
or space reflection. It means that the system possess strong symmetry
under rotation in charge space.

\end{document}